\documentclass[twocolumn,prb,showpacs]{revtex4}
\usepackage{graphicx}
\usepackage{dcolumn}
\usepackage{bm}

\begin{document}

\title {Wave Functions and Energies of Magnetopolarons in Semiconductor Quantum Wells}
\author{I. G. Lang, L. I. Korovin}
\address{A. F. Ioffe Physical-Technical Institute, Russian Academy
of Sciences, 194021 St. Petersburg, Russia}
\author{S. T. Pavlov\dag\ddag}
\address{\dag Facultad de Fisica de la UAZ, Apartado Postal C-580, 98060 Zacatecas, Zac., Mexico\\
\ddag P. N. Lebedev Physical Institute, Russian Academy of
Sciences, 119991 Moscow, Russia}

\begin {abstract}
The classification of magnetopolarons in semiconductor  quantum
wells (QW) is represented. Magnetopolarons appear due to the
Johnson - Larsen effect. The wave functions of usual and combined
magnetopolarons are obtained by the diodanalization of the
Schr\"odinger equation.
\end {abstract}


\maketitle

\section{Introduction}

The Johnson - Larsen effect  [1-3] arises at  a condition
\begin{equation}
\label{l} \omega_{LO}=j\omega_{e(h)H},
\end{equation}
where $ \omega _ {LO} $ is the frequency of a longitudinal optical
(LO) ôîíphonon,
$$\omega _ {e (h) H} = {|e|H\over cm _ {e (h)}} $$
is the cyclotron frequency, $m _ {e (h)} $ is the electron (hole)
 effective mass, $H $ is the magnetic field, $j $
is some number.\footnote {In a case of the "classical" Johnson -
Larsen effect  $j $ is the integer, but, as it will be shown
below, at some fractional values $j $ so-called "weak"
magnetopolaron effect arises.}

The Johnson - Larsen effect is colled also as a magnetopolaron
resonance, and the statets which are being formed under condition
 (1)  in semiconductors - by magnetopolarons. At
 magnetic fields appropriate to the  condition (1) the resonant connection between
the Landau bands with different quantum numbers $n $  (see Fig. 1)
arises. The electron-phonon interaction results into removing of a
degeneration in crossing points  of energy levels, what influences
magnetooptical effects. For the first time the magnetopolaron
states were discovered   in a bulk $InSb $ in the interband light
absorption [1-3].

After the pioneer Johnson-Larsen works the magnetopolaron effect
has attracted the  attention of  theoretical and experimental
groups. Magnetopolaron features in transport and  optical
phenomena were intensively investigated. During last years the new
wave of interest to the Johnson-Larsen effect was stimulated by
appearance of low-dimensional semiconductor objects, in which the
effect amplifies due to the size-quantization of electronic
excitations.

The formation of polaron states  takes place  in three-dimensional
(3D), and in quasi-two-dimensional (2D) systems. The distinction
between these systems consists in energy spectra of electrons
(holes)  at presence  of a quantizing magnetic field: in 3D
systems there appear one-dimensional Landau bands, in 2D -
discrete energy levels. This distinction results into different
splitting of  energy levels of an electron-phonon system.

In both 3D and 2D systems magnetopolaron states play an important
role in formation of frequency dependences of magnetooptical
effects, such as interband absorption of light, cyclotron
resonance and Raman scattering of light (see reviews [4-7]).

In [8] Korovin and Pavlov have shown that in bulk semiconductors
 the magnetopolaron splitting is proportional to $ \alpha ^ {2/3}
\hbar\omega _ {LO} $, where $ \alpha $ is the Fr\"ohlich
dimensionless electron-phonon coupling constant [9] ($ \alpha < <
1 $).

In quasi-2D systems (in particular in  semiconductor quantum wells
), effect amplifies, and the distance between components of
splitted  peaks (for example, of interband light absorption)
becomes proportional to $ \alpha ^ {1/2} \hbar\omega _ {LO} $
[10-17].

\section{Classification of magnetopolarons.}

In Fig. 1 the continuous lines represent terms of an
electron-phonon system concerning to the size-quantization quantum
numbers $l$. The model is used in which all phonons have the
disperionless frequency $ \omega _ {LO} $. On the abscissa axis
the relation $j ^ {-1} = \omega _ {e (h) H} /\omega _ {LO} $ is
represented, on the ordinate axis - the relation $E/\hbar\omega _
{LO} $, where $E $ is the energy counted from the energy $
\varepsilon_l ^ {e (h)} $, appropriate to $l $-th energy level of
the size-quantization (values $ \varepsilon_l ^ {e (h)} $ for QWs
of a finite depth are given, for example, in [18]).
\begin{figure}
\includegraphics[]{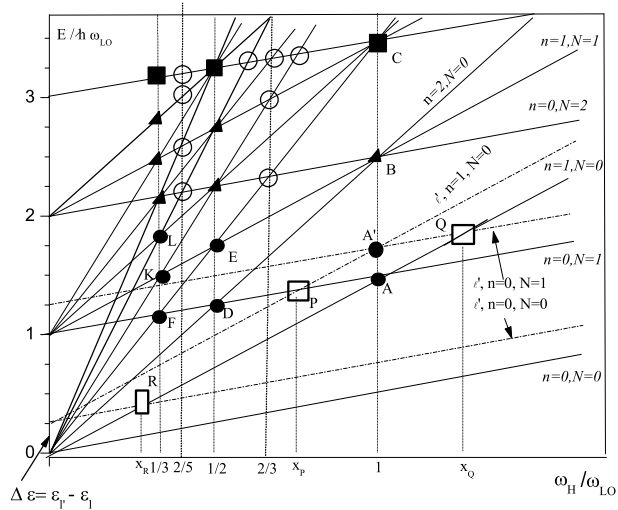}
\caption[*]{\label{Fig1.eps}  Energy levels of an electron (hole)
 - phonon system as functions of a magnetic
field. Points of crossings of lines correspond to polaron states.
Black circles are the  twofold polarons, triangles - threefold
polarons, squares - fourfold polarons. Empty circles are weak
polarons, empty squares - combined polarons, empty rectangular -
weak combined polaron. $E $ is the energy counted from energy $
\varepsilon_l $ of the size-quantization, $n $ is the Landau
quantum number, $N $ is the number of phonons; $l, l ^\prime $ are
the size-quantization quantum numbers.}
\end{figure}

The polaron states correspond to  crossing points of terms.  By
black circles "twofold" polarons are designated appropriate to
crossing only of two terms. Let us consider a certain crossing
point  to which number $j $ (see (1)) corresponds. Let $n $ is the
number of a Landau level crossing the given point  at $N=0 $. Then
for existence of a twofold polaron the condition to be carried
\begin{equation}
\label{2} 2j > n\geq j.
\end{equation}
It is easy to see that only one twofold polaron corresponds to
 $j=1 $ (designated by the
letter $A $). To  $j=2 $, i.e. $ \omega _ {H} /\omega _ {LO} =1/2
$, two twofold polarons ($ D $ and $E $) correspond, to $j=3 $,
i.e. $ \omega _ {H} /\omega _ {LO} =1/3 $, three twofold polarons
($F, K $ and $L $) correspond, etc.

In Fig. 1  polarons, located more to the left of $ \omega _ {H}
/\omega _ {LO} =1/3 $ are not shown. Let us notice that for
experiment the existence of polarons, appropriate to integers $ j
> 1 $ is extremely important. Really, the resonant $H_{res j} =
\omega_{LO} mc/j|e | $ decreases in $j $ times in comparison with
$H _ {res 1} $ for the polaron $A $.

Threefold polarons, appropriate to crossing of three terms, are
located above twofold polarons, higher -  fourfold polarons, etc.
In Fig. 1 threefold polarons are designated by black triangles,
fourfold - by black squares. The number of polarons of each grade
at given $j $ is equal to $j $. Threefold polarons in bulk
crystals for the first time were considered in [19], in QWs - in
[20-22].

Let us note that for crossing of three and more terms in one point
it is necessary an equidistance of Landau levels. In order the
theory [22] of
 threefold polarons would be applicable, it is necessary, that amendments to
 energies caused by non-parabolicity of a band or by an excitonic  effect [23],
 would be less than values of splittings of terms.
 But in case of twofold polarons the equidistance of levels
 is not an obstacle, as crossing of two terms  exists in any case.

 All above mentioned polarons correspond to the integer $j $.
 Besides in Fig. 1  other crossings of terms with
  quantum number $l $ (continuous lines) (designated by empty
  circles) are presented. They correspond to fractional  $j $. As
 the terms crossed in these points are characterized by values
 $ \Delta N\geq 2 $, real direct transitions between them with
 emission of one phonon are impossible. Let us name such polarons
 as weak. As the terms are crossed, their splitting is
 inevitably, but for calculation of value of splitting it is necessary to take into account
 transitions between crossed terms through virtual
 intermediate states or to take into account in the operator
   of electron-phonon interaction the small two-phonon contributions.
 As a result  $ \Delta E _ {weak} $ splittings of terms in case of weak
 polarons should be much smaller, than in a case of polarons
 for integer $j $. Contributions of transitions through intermediate states in
  $ \Delta E _ {weak} $ are of higher order on dimensionless
  constant  $ \alpha $, than $ \alpha ^ {1/2} $.

 At the account of two or more  of size-quantization quantum numbers  $l $
 the picture of crossing of terms
 becomes too complicated. Besides usual polarons, appropriate to a level
 $l ^\prime $ (for example, of the polaron $A ^\prime $), occur
 "combined" magnetopolarons, for which electron-phonon
  interaction connects two electron levels with different numbers
 $l $. Thus, the Landau numbers  can coincide or be different [24,
 25]. In Fig. 1 for an example   by strokes - dashed lines
 three terms concerning quantum number $l ^\prime $ are shown.
 Two combined polarons $P $ and $Q $ indicated by empty squares.
 In Fig. 1 it would be necessary to show more stroke - dotted
 lines and to obtain the greater number of combined polarons. However,
 it would strongly  complicate figure. For an example in Fig. 1
 polaron $R $ (light rectangular) is indicated, which is combined and
 weak.

 An interesting feature of combined polarons is that
  appropriate magnetic fields depend on
 distances $\Delta\varepsilon=\varepsilon_{l^\prime}-\varepsilon_l$
 between levels $l $ and
 $l ^\prime $  of size-quantization and,
 hence, on QW's depth and width. Really, with the help
 of Fig. 1 it is easy to obtain
\begin{eqnarray}
\label{3}
(\omega_H/\omega_{LO})_P=X_P=1-{\Delta\varepsilon\over\hbar\omega_{LO}},\nonumber\\
(\omega_H/\omega_{LO})_Q=X_Q=1+{\Delta\varepsilon\over\hbar\omega_{LO}}.
\end{eqnarray}

In Fig. 1 the case $ \Delta\varepsilon < \hbar\omega _ {LO} $ is
represented. If $ \Delta\varepsilon > \hbar\omega _ {LO} $, there
is only one combined polaron, to which there corresponds the
second  equality of (3). It is essential that the magnetopolaron
$P $ in Fig. 1 corresponds  to much smaller  $H $, than the
polaron $A $. It should facilitate its experimental observation.

One more kind combined polaron [2, 4] is not represented in Fig.
1, as it exists only under condition
\begin{equation}
\label{4} \Delta\varepsilon=\hbar\omega_{LO},
\end{equation}
when the terms $l ^\prime, n, N=0 $ and $l, n, N=1 $ coincide at
any magnetic fields. For performance of a resonant condition (4) a
certain distance between levels $l $ and $l ^\prime $ is required
, that is reached only by selection of QW width and depth.
Magnetic  field is required only for formation of Landau levels
and can be chosen rather small.  Under condition (4)"a special
polaron state " arises.

In order the picture represented in Fig. 1 woud be applicable, it
is necessary that the distances between the next levels $l, l-1,
l+1 $ were much greater, than $ \Delta E $ of polaron splittings.
As distance between levels decreases with the growth of QW width
$d $,  restrictions from above on $ d $ are imposed (numerical
estimations see in [26] (Fig. 2,3)).

\section{The Hamiltonian of a system.}

The energy  spectrum of  twofold magnetopolarons -  usual
(classical), and combined - was determined by two  ways giving
identical results. The first way was  used in [8] and it consists
in definition of poles of one-partical  Green function of an
electron. It also is applied in [25, 26]. Other way is described
in [18] and devoted to the polaron $A $. The polaron wave
functions
 are represented as a superposition of wave functions of the
unperturbed states (in a case of the polaron $A $ these are states
 $ (n=1, N=0) $
 and $ (n=0, N=1) $) with unknown  factors. The Schr\"odinger equation
 is reduced to a system of two equations for two factors.
 Equating determinant to 0, we obtain square-law equation for
 polaron energies of  states $p=a $ and $p=b $. Advantage in comparison with
  the first way is that simultaneously with the energy calculation
  we calculate the magnetopolaron wave functions. And these
  functions are necessary for the description of  many magnetooptical effects.

 In present work we generalize results of [18] for the polaron $A $ on
 case of any twofold polarons, including usual, combined and
 " special polaron state ". We pay the special attention  to
 wave functions, before unknown. The theory is not extended
 on weak, threefold, fourfold etc. polarons.

 Let us consider a semiconductor QW of type $I $ with the energy gap $E_g $
 and barrier $ \Delta E_e $ for electrons. For definiteness we
  investigate magnetopolarons with participation of electrons. Results
 can be easily used for description of magnetopolarons with
  participation of holes.

 The magnetic field is directed along an axis $z $ perpendicularly to the QW
 plane.
 The vector potential is chosen as $ {\bf A} = {\bf A} (0, xH,
 0) $. The Schr\"odinger equation  for electrons, interacting with $LO $
 phonons, looks like
\begin{equation}
\label{5} {\cal H}\Theta=E\Theta, {\cal H}={\cal H}_0+V, {\cal
H}_0={\cal H}_e+{\cal H}_{ph},
\end{equation}
and
\begin{equation}
\label{6} {\cal H}_e\Psi_{n, k_y,
l}=[(n+1/2)\hbar\omega_{eH}+\varepsilon_l]\Psi_{n, k_y, l},
\end{equation}
where
\begin{eqnarray}
\label{7}\Psi_{n, k_y, l}=\Phi_n(x+a_H^2k_y){1\over
\sqrt{L_y}}e^{ik_yy}\varphi_l(z),\nonumber\\
\Phi_n(x)={1\over\sqrt{\pi^{1/2}2^nn!a_H}}H_n(x/a_H)e^{-x^2/2a_H^2},
\end{eqnarray}
$a_H =\sqrt {c\hbar / | e|H} $ is the magnetic  length, $H_n (t) $
is the Hermitian polynomial. The functions $ \varphi_l (z) $ and
levels $ \varepsilon_l $ of energy  of size-quantization of
electrons in a QW  of finite depth are determined, for example, in
[18], $ {\cal H} _ {ph} $ is the Hamiltonian of the phonon system,
$V $ is the electron-phonon interaction.  In a case of an
indefinitely deep QW, when $\Delta E_e\rightarrow\infty$
$$\varphi_\ell(z)=\cases {\sqrt{2\over d}\quad sin({\pi
\ell z\over d} +{\pi \ell \over 2}), & $ |z|\leq {d\over 2}$, \cr
0 & $ |z|\geq {d\over 2}.$\cr},$$
$$\varepsilon_l(z)={\pi^2\hbar^2l^2\over 2m_ed},$$
$m_e $ is the electron effective mass.

Let us designate as $ \Psi _ {ph 0} (y) $ and $ \Psi _ {ph \nu}
(y) $ the wave functions of the phonon system appropriate to
absence of phonons and to presence of one phonon with indexes $
\nu\equiv ({\bf q} _ \perp, \mu) $, where $ {\bf q} _ \perp $is
the phonon wave vector  in a plane $xy, \mu $ are other indexes
[27], $Y $ are the coordinates of the phonon  subsystems. Let us
assume
\begin{equation}
\label{8} {\cal H}_{ph}\Psi_{ph 0}=0, {\cal H}_{ph}\Psi_{ph \nu
}=\hbar\omega_\nu\Psi_{ph \nu }.
\end{equation}

Functions with the large number of phonons  will be unnecessary,
as (it is visible in a Fig. 1) in formation of  classical twofold
polarons the appropriate states do not participate, the same as in
formation of combined polarons $P $ and $Q $.

We use model, in which the dispersion of $LO $ phonons  is not
taken into account, i.e. we believe
\begin{equation}
\label{9} \omega_\nu=\omega_{LO}.
\end{equation}
The influence of the phonon dispersion on  a magnetopolaron
spectrum is discussed in [23].

The electron-phonon interaction  looks like
\begin{equation}
\label{10} V=\sum_\nu[C_\nu({\bf r}_\perp, z)b_\nu+C_\nu^*({\bf
r}_\perp, z)b_\nu^+],
\end{equation}
where $b_\nu ^ + (b_\nu) $ is the phonon creation (annihilation)
operator,
\begin{equation}
\label{11}C_\nu({\bf r}_\perp, z)= C_\nu e^{i{\bf q}_\perp{\bf
r}_\perp}\xi_\nu(z),
\end{equation}
and $ \xi_\nu (z) $ is chosen so, that $ \xi_\nu (z=0) =1 $.

In a single QW instead of bulk $LO $ phonons there are three types
of phonons [27]. First, it is so-called  phonons semi-space, not
penetrating in a QW. Besides  interface phonons are available,
which damp at outside of QW. At last,  confined phonons exist in a
QW material. These fluctuations will not penetrate into a barrier,
their amplitude equals 0 on QW borders. In a case of confined
phonons [27] a set of indexes $ \nu $ includes $ {\bf q} _ \perp $
and discrete indexes $ \mu $, and the interaction (11) is
determined as
$$\xi_\nu(z)=\xi_\mu(z)=\cases { cos({\pi\mu z
 \over d}, \mu=1, 3\ldots,& $ |z|\leq
{d\over 2}$ \cr  sin({\pi\mu z
 \over d}, \mu=2, 4\ldots,& $ |z|\leq
{d\over 2}$ \cr 0 & $ \,|z|\geq {d\over 2}.$\cr},$$
\begin{equation}
\label{12}C_\nu=C_{{\bf q}_\perp, \mu}=
-\hbar\omega_{LO}\sqrt{{8\pi\alpha l\over
S_0d[q_\perp^2+(\mu\pi/d)^2]}},
\end{equation}
where $ \alpha $ is the Fr\"ohlich constant [9], $l =\sqrt
{\hbar/2m_e\omega _ {LO}}, S_0 $ is the normalization area.

In many theoretical calculations  of magnetopolaron spectra
 in a QW
the Fr\"ochlich interaction with  $LO $ phonons [9] is used. Thus,
$j=q_z, \xi_\nu (z) =e ^ {iq_zz}, $
\begin{eqnarray}
\label{13}C_\nu=C_{{\bf q}_\perp}=
-\hbar\omega_{LO}\sqrt{{4\pi\alpha l^3\over V_0}}{1\over
ql},\nonumber\\
\alpha={e^2\over2\hbar\omega_{LO}l}(\varepsilon_\infty^{-1}-\varepsilon_0^{-1}),
\end{eqnarray}

$V_0 $ is the normalization volume, $ \varepsilon_\infty
(\varepsilon_0) $ is the high-frequency (static) dielectric
permeability of QW.

In [26] it is investigated, when use of (13) is lawful for the
description of magnetopolaron spectra
 in a QW. It is shown that at a large sizes of
QW width  it is possible to neglect by interaction of electrons
with interface phonons, and the interaction  with confined phonons
(12) results in the same results as (13).

We do not concretize the form of the interaction (11) below.

\section{Wave functions and energies of magnetopolarons.}

Let us consider a polaron, arising due to crossing of terms $n_0,
l_0, N=1 $ and $n_1, l_1, N=0 $, where $n $ is the Landau quantum
number, $l $ is the quantum number of the size-quantization, $N $
is the number of phonons.  We search for the wave function as a
superposition
\begin{eqnarray}
\label{14} \Theta(x, y, z, Y)=\sum_{k_y}a_0(k_y)\Psi_{n_1, k_y,
l_1}(x, y, z)\psi_{ph 0}(Y)\nonumber\\
+\sum_{k_y, \nu}a_1(k_y)\Psi_{n_0, k_y, l_0}(x, y, z)\psi_{ph
\nu}(Y).
\end{eqnarray}
The indexes 0 and 1 at factors $a_0 (k_y) $ and $a_1 (k_y) $
designate the number of phonons. For convenience of the further
calculations we  introduce designations
\begin{eqnarray}
\label{15} \Psi_{n_1, k_y, l_1}(x, y, z)=\Psi_{1, k_y}(x, y,
z),\nonumber\\
\Psi_{n_0, k_y, l_0}(x, y, z)=\Psi_{0, k_y}(x, y, z)
\end{eqnarray}
and also
\begin{eqnarray}
\label{16}
\Sigma_1=(n_1+1/2)\hbar\omega_{eH}+\varepsilon_{l_1},\nonumber\\
\Sigma_0=(n_0+1/2)\hbar\omega_{eH}+\varepsilon_{l_0}.
\end{eqnarray}
Then the Schr\"odinger equation may be written down as
\begin{eqnarray}
\label{17} &&(E-\Sigma_1)\psi_{ph 0}\sum_{k_y}a_0(k_y)\Psi_{1, k_y}\nonumber\\
&+&(E-\Sigma_0-\hbar\omega_{LO})\sum_{k_y}a_1(k_y, \nu)\psi_{ph
\nu}\nonumber\\
&-& \psi_{ph 0}\sum_{k_y}\Psi_{0, k_y}\sum_\nu C_\nu({\bf
r}_\perp, z)a_1(k_y,
\nu)\nonumber\\
&-&\sum_{k_y}a_0(k_y)\Psi_{1, k_y}\sum_\nu C_\nu^*({\bf r}_\perp,
z)\psi_{ph \nu}=0.
\end{eqnarray}
In  (17) we used an approximation
$$ V\psi _ {ph \nu} (Y) \simeq C_\nu ({\bf r} _ \perp, z) \psi _ {ph 0} (Y), $$
because we consider the interaction only  between states with $N=0
$ and $N=1 $. All other possible transitions result in the
amendments of higher order on $ \alpha $.

Let us multiply (17)  on $ \psi _ {ph 0} ^ * (Y) $ and $ \psi _
{ph \nu ^\prime} ^ * (Y) $ and integrate on $Y $. Using  the
properties of the ortogonality   and normalization
 of phonon functions, we obtain two equations
\begin{eqnarray}
\label{18} &&(E-\Sigma_1)\sum_{k_y}a_0(k_y)\Psi_{1,
k_y}\nonumber\\
&-&\sum_{k_y}\Psi_{0, k_y}\sum_\nu C_\nu({\bf r}_\perp, z)a_1(k_y,
\nu)=0,\nonumber\\
&&(E-\Sigma_0-\hbar\omega_{LO})\sum_{k_y}a_1(k_y, \nu)\Psi_{0,
k_y}\nonumber\\
&-&\sum_{k_y}\Psi_{1, k_y} C_\nu^*({\bf r}_\perp, z)a_0(k_y)=0.
\end{eqnarray}
First of the equations (18) we multiply on $ \Psi _ {1, k_y
^\prime} ^ * (x, y, z) $, second - on $ \Psi _ {0, k_y ^\prime} ^
* (x, y, z) $ and we integrate on $x, y, z $. We obtain
\begin{eqnarray}
\label{19} (E-\Sigma_1)\sum_{k_y}a_0(k_y)\delta_{k_y,
k_y^\prime}\nonumber\\- \sum_{k_y, \nu}a_1(k_y, \nu)M^*(k_y,
k_y^\prime, \nu)=0,\nonumber\\
(E-\Sigma_0-\hbar\omega_{LO})\sum_{k_y}a_1(k_y, \nu)\delta_{k_y,
k_y^\prime}\nonumber\\- \sum_{k_y}a_0(k_y)M(k_y, k_y^\prime,
\nu)=0,
\end{eqnarray}
where the designation for a matrix element is introduced
\begin{eqnarray}
\label{20}M(k_y, k_y^\prime, \nu)=\int dxdydz \Psi_{0,
k_y^\prime}^*(x, y, z)\nonumber\\
\times C^*_\nu({\bf r}_\perp, z)\Psi_{1, k_y}(x, y, z).
\end{eqnarray}
Using designations (11) and (15) we obtain
\begin{eqnarray}
\label{21}M(k_y, k_y^\prime, \nu)=\delta_{k_y-q_y,
k_y^\prime}U^*(\nu)\nonumber\\
\times e^{ia_H^2q_x(k_y^\prime+q_y/2)},
\end{eqnarray}
where
\begin{equation}
\label{22}U^*(\nu)=C_\nu^*{\cal K}_{n_1n_0}(a_Hq_y-a_Hq_x){\cal
M}^*(\nu),
\end{equation}
\begin{eqnarray}
\label{23}&&{\cal K}_{nm}(p_x, p_y)={\cal K}_{nm}({\bf
p})=\left[{min(n!, m!)\over max(n!, m!)}\right]^{1/2}\nonumber\\
&\times& i^{|n-m|}\left({p\over \sqrt
{2}}\right)^{|n-m|}e^{-p^2/4}e^{i(\phi-\pi/2)(n-m)}\nonumber\\
&\times& L^{|n-m|}_{min(n, m)}(p^2/2),
\end{eqnarray}
$p =\sqrt {p_x^2+p_y^2}, \phi =\arctan (p_y/p_x), L^n_m (t) $ is
the Laguerre polynomial,
\begin{equation}
\label{24}{\cal M}(\nu)=\int dz
\varphi_{l0}(z)\varphi_{l1}(z)\xi_\nu(z).
\end{equation}
In (22) we used the integral
\begin{equation}
\label{25}{\cal K}_{nm}(x, y)=e^{ixy/2}\int dt
f_m(t)f_n(t+x)e^{ity},
\end{equation}
where
 $$f_n(t)={1\over\sqrt{\sqrt{2}~2^nn!}}e^{-t^2/2}H_n(t).$$
Having substituted (21) in (18) and having executed summation on
$k_y $, we obtain
\begin{eqnarray}
\label{26} &&(E-\Sigma_1)a_0(k_y)- \sum_{\nu}a_1(k_y- q_y,
\nu)\nonumber\\
&\times& e^{-ia_H^2q_x(k_y-q_y)/2}U(\nu)=0,\nonumber\\
&&(E-\Sigma_0-\hbar\omega_{LO})a_1(k_y, \nu)-a_0(k_y+q_y)\nonumber\\
&\times& e^{-ia_H^2q_x(k_y+q_y)/2}U^*(\nu)=0.
\end{eqnarray}
We have from the second equation
\begin{eqnarray}
\label{27} a_1(k_y,
\nu)&=&a_0(k_y+q_y)e^{ia_H^2q_x(k_y+q_y)/2}\nonumber\\
&\times&{U^*(\nu)\over E-\Sigma_0-\hbar\omega_{LO}}.
\end{eqnarray}
Having substituted (27) in the first equation, we obtain the
square-law equation for energy $E $
\begin{eqnarray}
\label{28} (E-\Sigma_1)(E-\Sigma_0-\hbar\omega_{LO})\nonumber\\
\times\sum_\nu |U(\nu)|^2=0.
\end{eqnarray}
Let us introduce a designation
\begin{equation}
\label{29}w(n_o, n_1, l_0, l_1)= \sum_\nu|U(\nu)|^2.
\end{equation}
Then it follows  from (22)
\begin{eqnarray}
\label{30}&&w(n_o, n_1, l_0, l_1)=\nonumber\\
&=&\sum_\nu |C_\nu|^2B_{n_0n_1}(a^2_Hq_\perp^2/2)|{\cal M}_{l_0,
l_1}(\nu)|^2,
\end{eqnarray}
where
\begin{eqnarray}
\label{31}B_{n_0n_1}(u)={min(n_0!, n_1!)\over max(n_0!, n_1!)}u^{|n_0-n_1|}\nonumber\\
\times e^{-u} [L^{|n_0-n_1|}_{min(n_0, n_1)}(u)]^2.
\end{eqnarray}

The equation (28) has two solutions
\begin{eqnarray}
\label{32}E_{a, b}={1\over
2}\{\Sigma_0+\Sigma_1+\hbar\omega_{LO}\nonumber\\
\pm\sqrt{(\Sigma_1-\Sigma_0-\hbar\omega_{LO})^2+4w(n_o, n_1, l_0,
l_1)}\},
\end{eqnarray}
where the indexes $a $ and $b $ correspond to  $ + $ and $ - $.
The energy distance between two ìàãímagnetopolaron states is equal
\begin{equation}
\label{33}\Delta E=\sqrt{\lambda^2+4w(n_o, n_1, l_0, l_1)}\},
\end{equation}
where
$$\lambda=(n_1-n_0)\hbar\omega_{eH}-\hbar\omega_{LO}+\varepsilon_{l_1}-\varepsilon_{l_2}$$
describes a deviation from an exact resonance. An energy spectrum
of anyone twofold polaron (classical or combined) is schematically
represented in Fig. 2.
\begin{figure}
\includegraphics[]{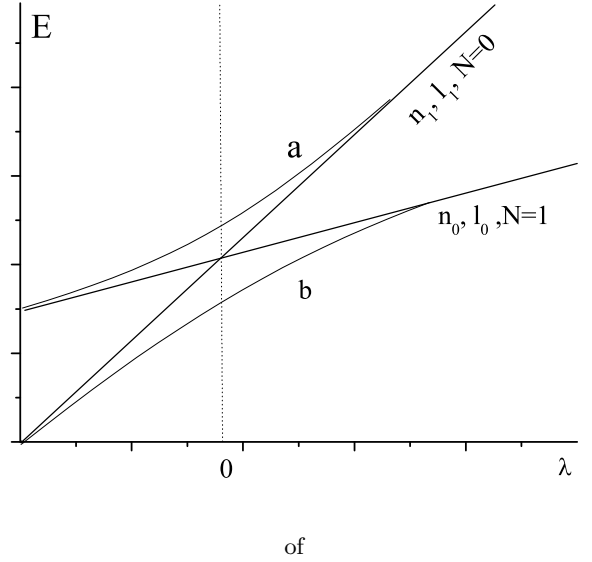} of 
 \caption[*]{\label{Fig2.eps}The schematical image of polaron
 energies
 formed on crossing $a $ and $b $. On an ordinate
axis  the value
 $\lambda=\Sigma_1-\Sigma_0-\hbar\omega_{LO}$, where $
\Sigma_1 = (n+1/2) \hbar\omega _ {eH} + \varepsilon _ {l_1},
\Sigma_0 = (n+1/2) \hbar\omega _ {eH} + \varepsilon _ {l_0} $
 is represented.}
\end{figure}

Using (14), (27), (32) and designation (15), we obtain the
magnetopolaron wave functions  for states $p=a $ and $p=b $
\begin{eqnarray}
\label{34}&&\Theta_p(x, y, z,
Y)=\sum_{k_y}a_{0p}(k_y)\Big[\Psi_{1,
k_y}(x, y, z)\psi_{ph 0}(Y)\nonumber\\
&+&(E_p-\Sigma_0-\hbar\omega_{LO})^{-1}\sum_\nu\exp[ia_H^2q_x(k_y-q_y/2)]\nonumber\\
&\times& U^*_\nu\Psi_{0, k_y-q_y}(x, y, z)\psi_{ph \nu}(Y)\Big].
\end{eqnarray}
The direct calculation shows that functions with indexes $p=a $
and $p=b $ are ortogonal, i.e.
\begin{equation}
\label{35}\int dY d^3r \Theta^*_b\Theta_a=0,
\end{equation}
and from the normalizaion condition
\begin{equation}
\label{36}\int dY d^3r \Theta^*_p\Theta_p=1
\end{equation}
we obtain the requirement
\begin{equation}
\label{37}\sum_{k_y}|a_{0p}(k_y)|^2=\left[1+{w(n_0, n_1, l_0,
l_1)\over (E_p-\Sigma_0-\hbar\omega_{LO})^2}\right]^{-1}.
\end{equation}

Let us choose factors $a _ {0p} (k_y) $ as
\begin{equation}
\label{38}a_{0p}(k_y)=\delta_{k_y, k_y^\prime}\left[1+{w(n_0, n_1,
l_0, l_1)\over (E_p-\Sigma_0-\hbar\omega_{LO})^2}\right]^{-1/2}.
\end{equation}
Then the polaron wave functions  are characterized by indexes $p $
and $k_y $ in designations (15) and  finally
\begin{eqnarray}
\label{39}&&\Theta_{p, k_y}(x, y, z, Y)=\left[1+{w(n_0, n_1, l_0,
l_1)\over
(E_p-\Sigma_0-\hbar\omega_{LO})^2}\right]^{-1/2}\nonumber\\
&\times&[ \Psi_{1, k_y}(x, y, z)\psi_{ph
0}(Y)+(E_p-\Sigma_0-\hbar\omega_{LO})^{-1}\nonumber\\
&\times&\sum_\nu\exp[ia_Hq_x(k_y-q_y/2)]U^*(\nu)\nonumber\\
&\times&\Psi_{0, k_y-q_y}(x, y, z)\psi_{ph \nu}(Y)].
\end{eqnarray}
The conditions of ortogonality and normalization are carried out
\begin{equation}
\label{40}\int dY d^3r \Theta_{p^\prime, k_y^\prime}^*\Theta_{p,
k_y}=\delta_{p, p^\prime}\delta_{k_y, k_y^\prime}.
\end{equation}
With the help of wave functions (39) we determine the probability
to find a system in a state  with zero phonons and with one $LO $
phonon with an index $ \nu $. We have
\begin{eqnarray}
\label{41}Q_{0p}&=&\left[1+{w\over(E_p-\Sigma_0-\hbar\omega_{LO})^2}\right]^{-1}\nonumber\\
&=&{1\over 2}\left(1\pm{\lambda\over\sqrt{\lambda^2+4w^2}}\right),
\end{eqnarray}
\begin{eqnarray}
\label{42}Q_{\nu p}&=&{|U(\nu)|^2\over(E_p-\Sigma_0-\hbar\omega_{LO})^2}\nonumber\\
&\times&\left[1+{w\over(E_p-\Sigma_0-\hbar\omega_{LO})^2}\right]^{-1}.
\end{eqnarray}
Summarizing $Q _ {\nu p} $ on $ \nu $ we obtain the total
probability to find our system in a state with one phonon
\begin{eqnarray}
\label{43}Q_{1 p}&=&\sum_\nu Q_{\nu p}={1\over
2}\left(1\mp{\lambda\over\sqrt{\lambda^2+4w^2}}\right)\nonumber\\
&=&1-Q_{0 p}.
\end{eqnarray}
In all formulas the top sign corresponds to $p=a $, and bottom -
to $p=b $. In  an exact resonance, when $ \lambda=0 $ or
\begin{equation}
\label{44}\hbar\omega_{eH}n_1+\varepsilon_{l_1}=\hbar\omega_{eH}n_0+\varepsilon_{l_1}+\hbar\omega_{LO},
\end{equation}
which is reached at the resonant  $H _ {res} $ magnetic field, the
energies of polaron states  are equal
\begin{equation}
\label{45}E_{a, b}^{res}=\Sigma_1\pm\sqrt{w(n_0, n_1, l_0, l_1)},
\end{equation}
and the polaron splitting is as follows
\begin{equation}
\label{46}\Delta E^{res}=2\sqrt{w(n_0, n_1, l_0, l_1)}.
\end{equation}
Numerical calculations of  $ \Delta E ^ {res} $ for some polarons
are given in [25, 26].

In the exact resonance  probabilities of states without phonons
and with one phonon are equal, i. e.
\begin{equation}
\label{47}Q_{0 p}=Q_{1 p}=1/2.
\end{equation}

Let us consider a situation far away from the resonance, when
\begin{equation}
\label{48}|\Sigma_1-\Sigma_0-\hbar\omega_{LO}|>>\Delta E^{res}.
\end{equation}
The results are different for cases $
\Sigma_1-\Sigma_0-\hbar\omega _ {LO} < 0 $ and $
\Sigma_1-\Sigma_0-\hbar\omega _ {LO} > 0 $. In Fig. 2
 the left part from a point $
\Sigma_1-\Sigma_0-\hbar\omega _ {LO} =0 $ corresponds to the first
case, right - to the second case. Introducing the indexes $left $
and $right $ we obtain
\begin{equation}
\label{49}E_{a right}=E_{b left}=\Sigma_1+{w\over
\Sigma_1-\Sigma_0-\hbar\omega_{LO}},
\end{equation}
\begin{equation}
\label{50}E_{a left}=E_{b right}=\Sigma_0+\hbar\omega_{LO}+{w\over
\Sigma_0-\Sigma_1+\hbar\omega_{LO}},
\end{equation}
\begin{eqnarray}
\label{51}&&\Theta_{a, k_y,right}=\Theta_{b, k_y,left}=\Psi_{1,
k_y}\psi_{ph 0}\nonumber\\
&+&{1\over
\Sigma_1-\Sigma_0-\hbar\omega_{LO}}\sum_\nu\exp[ia_Hq_x(k_y-q_y/2)]
\nonumber\\
&\times& U^*(\nu)\Psi_{0, k_y-q_y}\psi_{ph \nu},
\end{eqnarray}
\begin{eqnarray}
\label{52}&&\Theta_{a, k_y,left}={\sqrt{w}\over
\Sigma_0-\Sigma_1+\hbar\omega_{LO}}\Psi_{1, k_y}\psi_{ph
0}\nonumber\\
&+&{1\over\sqrt{w}}
\sum_\nu\exp[ia_Hq_x(k_y-q_y/2)]U^*(\nu)\nonumber\\
&\times&\Psi_{0, k_y-q_y}\psi_{ph \nu},
\end{eqnarray}
\begin{equation}
\label{53}\Theta_{b, k_y,right}=-\Theta_{a, k_y,left},
\end{equation}

\footnote {In [18] the misprint is admitted: in the right part of
the analogue of our formula (53)  $ + $ is put instead of  $ - $
.}

The results (49) - (53) are in accordance with formulas of the
perturbation theory (see, for example, [28, page 165]) if to take
into account only two states of the system with indexes $n_0, l_0,
N=1 $ and $n_1, l_1, N=0 $. Amendments to the energy are
proportional $ \alpha $, amendments to the wave functions are
proportional to $ \alpha^{1/2} $. Far away from the resonance in a
point $ \Sigma_1 =\Sigma_0 +\hbar\omega _ {LO} $ we have to take
into account possible transitions in other states of our system.

Thus, the energy spectra and wave functions of usual (classical)
  and combined magnetopolarons in semiconductor QWs
are calculated. These functions are necessary for theoretical
consideration of the optical phenomena, in which the Johnson -
Larsen effect  is essential.

\end{document}